The detection of acute kidney injury with hyperpolarized $^{13}$C Urea and multi-exponential fitting.

James T. Grist1, Christian Østergaard Mariager2, Haiyun Qi2, Per Mose Nielsen2, Christoffer Laustsen2.

1 The Institute of Child Health, Institute of Cancer and Genomic Sciences, School of Medical and Dental Sciences, The University of Birmingham, Birmingham, UK.

2 MR Research Centre, Department of Clinical Medicine, Aarhus University, Aarhus, Denmark.

**Abstract**

**Purpose:** To assess the utility of Laplacian fitting to describe the differences in hyperpolarized $^{13}$C urea $T_2$ relaxation in ischemic and healthy rodent kidneys.

**Theory and Methods:** Six rats with unilateral renal ischemia were investigated. $^{13}$C urea $T_2$ mapping was undertaken with a radial fast spin echo method, with subsequent post-processing performed with regularised Laplacian fitting.

**Results:** Simulations showed that Laplacian fitting was stable down to a signal to noise ratio of 20. *In vivo* results showed a significant increase in the mono- and decrease in bi-exponential pools in IRI kidneys, in comparison to healthy (14±10% vs 4±2%, 85±10% vs 95±3%, p<0.05).

**Conclusion:** We demonstrate, for the first time, the differences in multi-exponential behaviour of $^{13}$C,$^{15}$N$_2$-urea between the healthy and ischemic rodent kidney. The distribution of relaxation pools were found to be both visually and numerically significantly different. The ability to improve the information level in hyperpolarized MR, by utilizing the relaxation contrast mechanisms is an appealing option, that can easily be adopted in large animals and even in clinical studies in the near future.

**Introduction**

Hyperpolarized [13]C MRI, recently translated in to clinical studies and offers the exciting possibility to non-invasively map biochemical processes and enzymatic activity(1–4). Several novel imaging biomarkers have been demonstrated using a variety of hyperpolarized imaging probes beyond hyperpolarized [1-[13]C]pyruvate, such as tissue necrosis using [1,4-[13]C]fumarate, pH with [13]C-bicarbonate and perfusion with urea(5–7) . [13]C,[14]N$_2$-urea and the longer lived [13]C,[15]N$_2$-urea has shown great promise for renal examinations for assessing the perfusion and renal functional dependent urea distribution, in a number of renal physiological and pathophysiological models (7–15).

Previous results showed a correlation between renal oxygenation, and [1]H/[13]C urea relaxation, providing a promising surrogate for measuring tissue oxygenation (9). Further to the single-exponential method used in a previous study, a study focusing on the bi-exponential relaxation of [13]C[15]N$_2$ urea has been undertaken – showing both the large increase in signal to noise achieved with a [15]N substitution and the bi-exponential relaxation of urea in the corticomedullary gradient, with the long and short components attributed to urea in the collecting ducts and renal interstitium, respectively(16). Further to a bi-exponential model, it is possible to quantify relaxation with a multi-exponential Laplacian model(9). This method has been demonstrated in a number of studies in both [1]H and [23]Na MRI, however never applied to *in vivo* [13]C relaxation modelling in either healthy or acute studies(17,18). The advantage of using a Laplacian approach to relaxation modelling is that it is possible to ascertain if single- or multi-exponential relaxation is exhibited in a given voxel by determining the most likely distribution of T$_2$ components.

Laplacian fitting is performed by decomposing multi-echo data into discrete relaxation (here T$_2$) components through an inverse Laplacian transform as shown in equation 1.1.

$$z(TE) = \sum_1^k g(\mathrm{k})\exp\left(-\frac{TE}{T_{2,k}}\right) \text{ 1.1}$$

Where z(t) is the time domain signal at each echo time (TE), $g(\mathrm{k})$ is the relaxation distribution, and $T_{2,k}$ are the relaxation distribution components. This technique is inherently challenging due to the non-unique solutions produced from standard fitting methods, however can be appropriately performed with regularisation and physical constraints on the distribution of $g(\mathrm{k})$ (17,19).

**Methods**

**Simulations**

Bi-exponential data were simulated in Matlab (2018b, The Mathworks, MA) using equation 1.1.

$$Signal = a \exp\left(-\frac{Time}{T_2^1}\right) + b \exp\left(-\frac{Time}{T_2^2}\right) \; 1.1$$

The pool sizes, a and b, were kept constant at 0.7 and 0.3, with relaxation constants of 100 and 1000ms, respectively. Multiple data sets were simulated with varying signal to noise ratios (SNR) from 100 to 5, and 64 echoes with varying temporal resolution (10:50ms). Bland Altmann analysis was performed comparing the calculated pool size and relaxation constants with ground truth.

**Rodent protocol**

The study is a post hoc study, using previously acquired data on six male Wistar rats(9). All rats were subjected to unilateral renal ischemia by clamping the left renal artery with a non-traumatic clamp for 40min and a reperfusion period of 24h, similar to a previously reported procedure(20). Immediately after the scan session, the animals were killed under the anaesthesia. Temperature and respiration were monitored during both the surgical procedure and the MRI scan session.

**Hyperpolarization**

Hyperpolarized 145mM urea samples were prepared by adding 200mL [$^{13}$C,$^{15}$N$_2$]urea (Sigma-Aldrich, Broendby,DK), glycerol (Sigma-Aldrich, Broendby, DK) and AH111501 (GE Healthcare, Broendby, DK) (6.4M concentration) mixed ratio (0.30:0.68:0.02, respectively) to a fluidpath (GE Healthcare, Broendby, DK) and placing it in the 5T SPINlab polarizer (GE Healthcare, Broendby, DK). Samples were polarized for 3 hours and then rapidly dissolved and transferred to the rats placed in a 9.4T preclinical MR scanner (Agilent, UK) equipped with a $^1$H/$^{13}$C dual-tuned volume coil (Doty scientific, Columbia, SC). Injection volume was approximately 1.0mL.

**$^{13}$C-T$_2$ MRI**

Hyperpolarized $^{13}$C-urea T$_2$ mapping was performed with single shot 2D golden-angle radial fast spin echo (repetition time = 3000 ms, echo time (TE) = 4.6 ms, ΔTE=36.8 ms, number of echoes = 64, field of view = 70x70mm$^2$, matrix = 64x64  flip angle 90/180 degrees for the slice selective excitation/refocusing pulses, 10mm slice thickness(21)). Reconstruction of the radial data was performed in Gadgetron(22).

Post-Processing

**Signal to Noise ratio**

Signal to noise ratio maps were produce by dividing each voxel in the first echo image by the signal standard deviation in a region outside of the rodent in the final echo image (assumed to be noise). Regions with SNR less than 20 were excluded from fitting.

**Laplacian fitting**

Laplacian fitting was performed on both simulated and experimental data using a regularized least-squares approach implemented in Matlab, with no prior assumptions on the initial relaxation distribution, constrained to exist between 10 and 3000ms (23).

Fitting was halted when either the $Chi^2$ of the time domain fit was greater than 98%, or an iteration limit was reached. If $Chi^2$ was not above 98% after reaching the iteration limit, the voxel was discarded.

**Regions of interest**

Regions of interest were drawn in Matlab to segment each kidney in the axial plane to produce both healthy and ischemic kidney masks.

**Analysis**

After Laplacian transform, the number of pools in the multi-exponential relaxation distribution was determined using an automated Gaussian fit to each relaxation peak. The number of voxels in an IRI or healthy kidney mask exhibiting a mono- and bi- exponential decay was calculated over all subjects and averaged.

Differences in the % of mono-, and bi- exponential voxels between healthy and IRI kidneys was assessed using a shapiro-wilk test and either a Mann-Wintney U test or unpaired t-test if normally distributed, assuming significance at p<0.05.

## Results

**Laplacian fitting is stable to low signal to noise ratios.**

Laplacian fitting was found to be stable between echo spacings of 10 ms to 90 ms (Figure 1A and B) and from SNR 100 to 20 (Bland-Altmann plots shown in Figure 1B and C). Below SNR 20 there was a rapid drop off in fit performance, with poor results observed in comparison to ground truth pool size and relaxation time constant values.

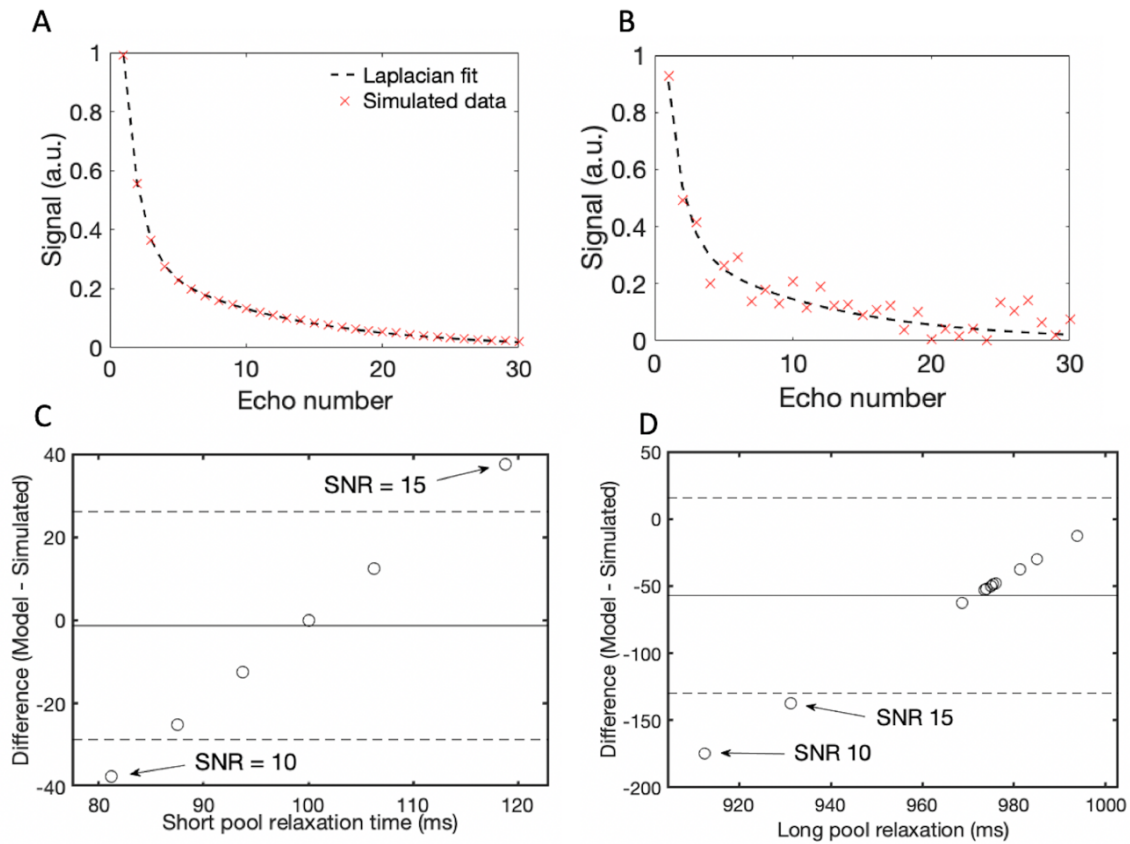

**Figure 1** - Simulation results showing Laplacian fitting with time resolution 10ms (A) and 90ms (B) were stable to SNR 20 as shown in Bland-Altmann plots in C (short pool relaxation time) and D (long pool relaxation time), only odd echoes in A and B plotted for ease of visualization.

**Laplacian fitting reveals multi-exponential relaxation of hyperpoalrized urea in the IRI and healthy kidney.**

Imaging was successful in all rats, with example [1]H and [13]C imaging data shown in figure 2 A and B, respectively. The average SNR in the healthy and IRI kidneys remained above 20 for the first 40 echoes, with the final echoes decaying to SNR < 3 (assumed to be the noise floor). Fitting results showed two relaxation pools in both the healthy and IRI kidney, resulting in clear multi-exponential behaviour. An example fit showing a comparison of single and Laplacian fitting in a healthy kidney voxel is shown in figure 3 with $Chi^2$ for single exponential and Laplacian multi-exponential fit are 0.58 and 0.99 respectively.

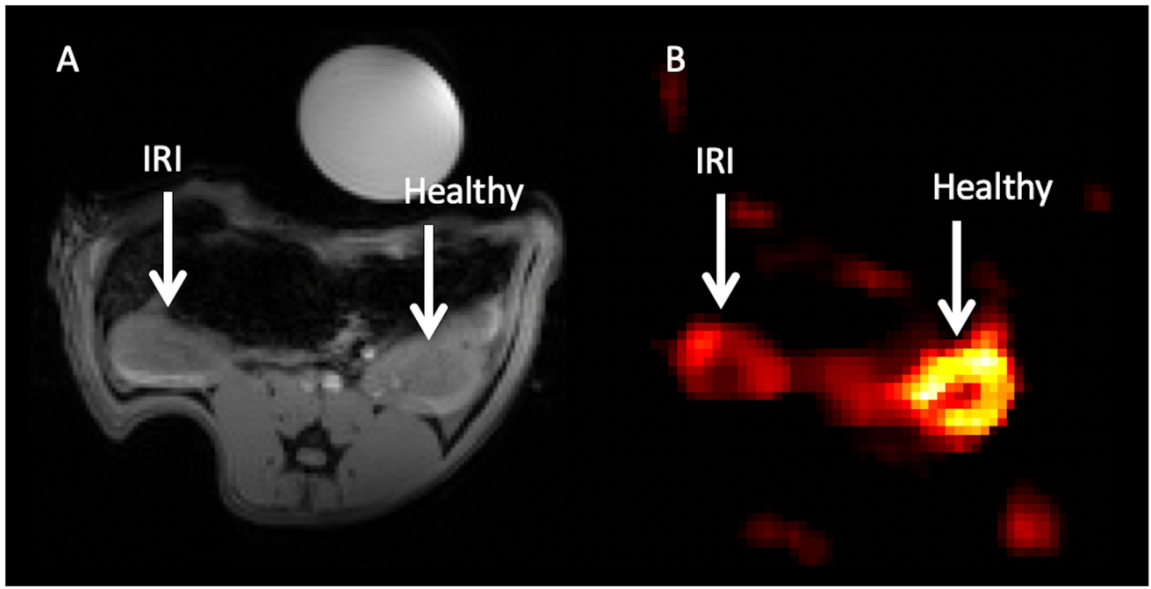

**Figure 2** – Example proton (A) and $^{13}C$ Urea (B) imaging of healthy and IRI kidneys.

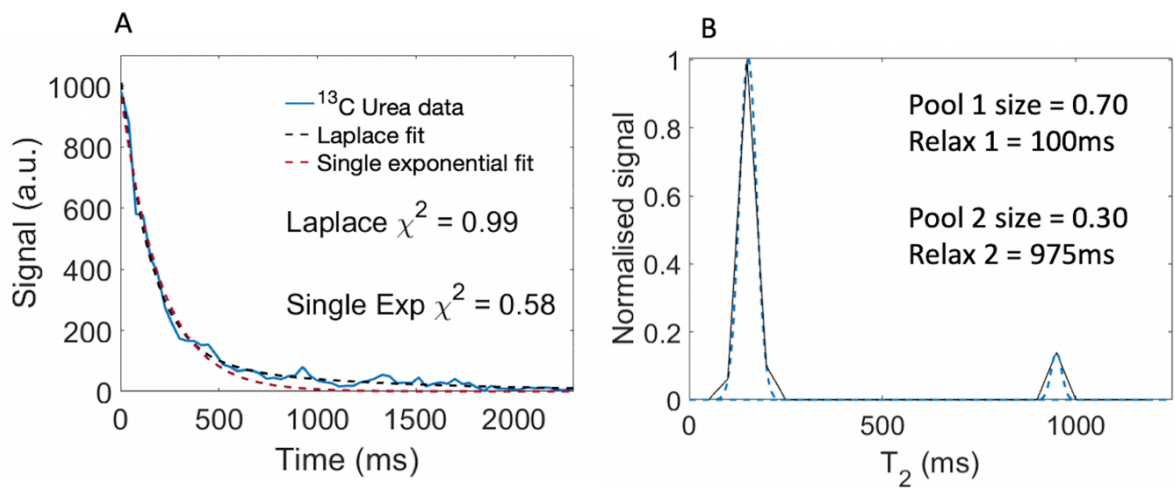

**Figure 3** – Example fitting of urea relaxation data (black, dotted) from a healthy kidney voxel with a single-exponential (red, dotted) and improved Chi2 with Laplacian (blue) fitting. Example bi- (B, distribution cut at 1500ms for ease of viewing) exponential relaxation pool from a healthy kidney voxel. The blue dotted line represents the Gaussian fits applied during post-processing.

**AKI causes a decrease in bi-exponential behaviour in the injured renal system.**

Statistical analysis revealed a marked difference in the number of voxels exhibiting mono- and bi-exponential relaxation behaviour between the IRI and healthy kidneys (14±10% vs 4±2%, 85±10% vs 95±3%, respectively, p < 0.05). Further to the statistical significance, a clear visual injury is seen in the IRI kidney relaxation pools map (see figure 4 A, and B for imaging and C for pool size results) allowing for a visual delineation of the injury.

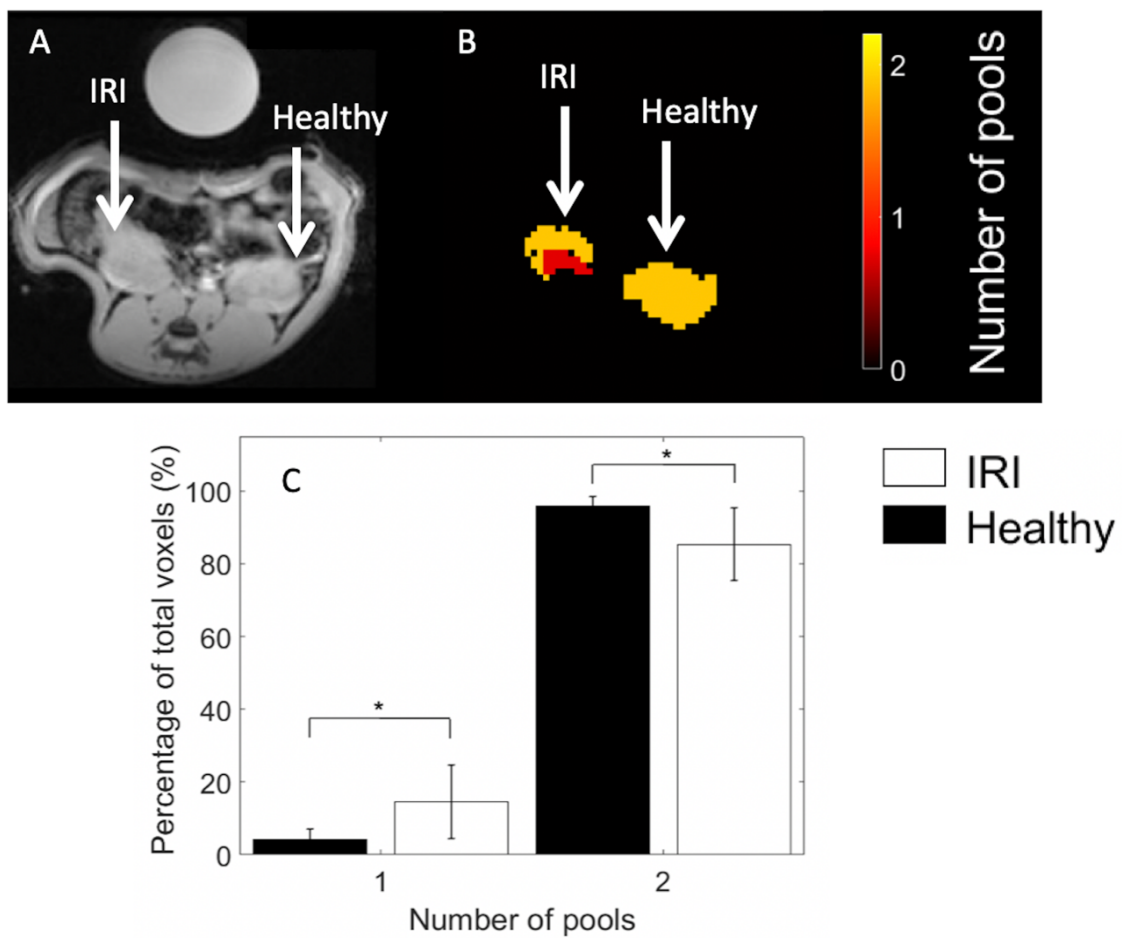

**Figure 4** – Example 1H (A) and pool maps of IRI and healthy kidneys (B) and quantitative analysis of pool distribution in healthy (white) and IRI (black) kidneys. (C) A significant difference between the IRI and healthy kidney between the mono- and bi-exponential pools (14+-10 vs 4+-2% , 85+-10 vs 95+-3 % , respectively, * = p < 0.05 in both cases).

**Discussion**

Results from this study demonstrate the power of hyperpolarized $^{13}$C-urea in detecting abnormalities associated with renal damage, with visible derangement in the relaxation distribution across the IRI kidney, relative to healthy. The alteration in relaxation pools could originate from decreased perfusion in the kidney, found in previous studies[20], as well as reduced water reabsorption leading to an increased blood pool fraction for urea relaxation. Indeed, we speculate that the mono-exponential pool is largely determined by the blood pool, and the bi-exponential the combined blood and tissue pools in the renal system. The lack of two pools in the medulla and pelvic regions of the IRI kidney suggest the reduced concentration ability of the IRI kidney to exclude urea uptake, and thus exhibit a largely extracellular single component urea pool in the inner parts of the kidney[7,24–26].

The ability to directly image the effects of acute kidney injury (AKI) with magnetic resonance imaging (MRI) or computed tomography (CT) is increasingly hampered by the cautioned use of gadolinium containing contrast agents to map renal perfusion or the radiation exposure, respectively[27]. Therefore, an imaging method that can overcome these challenges is highly desirable. Due to the rapid nature of the hyperpolarized imaging session, combined with an acquisition that is relatively motion insensitive, this method could provide a very powerful tool for clinical studies. Indeed, the technique may be able to estimate Glomerular filtration rate, perfusion, and potentially even oxygenation in one single scan[28].

Further improvements to this study could be found through utilising multi-slice imaging, to evaluate the injury in the whole kidney, providing a more accurate spatial localisation of the injury. It is also paramount that these results are translated to large animal or clinical studies, to prove the concept on a clinical system.

**Conclusion**

In conclusion, this study has assessed and demonstrated the utility of multi-exponential fitting in assessing the relaxation changes of hyperpolarized urea in acute kidney injury. Here results showed that the injury was detectable using simple masking results assessing the mono- and bi-exponential relaxation distribution in healthy and IRI kidneys.

## Acknowledgements

The authors would like to thank Henrik Vestergaard Nielsen for his laboratory assistance.